\documentclass[11pt,a4paper]{article}
\usepackage[margin=25mm]{geometry}
\usepackage[english]{babel}
\usepackage[utf8]{inputenc}
\usepackage{amsmath}
\usepackage{graphicx}
\usepackage[colorinlistoftodos]{todonotes}
\usepackage{amsmath}
\usepackage{amssymb}
\usepackage{yfonts}
\usepackage{amsmath}
\usepackage{verbatim}
\usepackage[T1]{fontenc}
\usepackage[utf8]{inputenc}
\usepackage{authblk}

\providecommand{\keywords}[1]
{
  \small	
  \textbf{\textit{Keywords---}} #1
}
 
\title{Correspondence Between the Energy Equipartition Theorem in Classical Mechanics and its Phase-Space Formulation in Quantum Mechanics}

\author[1]{Esteban Marulanda\thanks{esteban.marulandaa@udea.edu.co}}
\author[1]{Alejandro Restrepo
\thanks{alejandro.restrepo19@udea.edu.co}}
\author[1,2]{Johans Restrepo\thanks{johans.restrepo@udea.edu.co}}
\affil[1]{Instituto de Física, Universidad de Antioquia, Calle 70 No. 52-21, Medellín, Colombia}
\affil[2]{Group pf Magnetism and Simulation. Instituto de Física, Universidad de Antioquia. A.A. 1226, Medellín, Colombia}
\date{\today}

\begin{document}
\maketitle
\begin{abstract}
\noindent In classical physics there is a well-known theorem in which it is established that the energy per degree of freedom is the same. However, in quantum mechanics due to the non-commutativity of some pairs of observables and the possibility of having non-Markovian dynamics, the energy is not equally distributed. We propose a correspondence between what we know about the classical energy equipartition theorem and its possible counterpart in phase-space formulation in quantum mechanics based on the Wigner representation. Also, we show that in the high-temperature regime, the classical result is recovered.

\end{abstract}

\keywords{Energy equipartition theorem, Phase-space quantum mechanics }

\section{Introduction}

The energy equipartition theorem is one of the most important results in the classical theory of statistical mechanics due to its quantitative predictions and applicability in many areas of physics \cite{Greiner1995}. However, few references exist for an extension of it when quantum effects become relevant, referring mainly to low-temperature phenomena \cite{lowtemp}. 
A better understanding of the energetic distributions in this regime is then necessary for progressing in theoretical aspects and applications of quantum systems. 
\vspace{1.2mm}

\noindent Currently, the authors working in the area agree that the equipartition of energy no longer holds in the quantum regime and the energetic distribution follows a better-called energy partition theorem supported in the construction of a distribution function \cite{art1,art2,art3}. Applications to a few models have been made with satisfactory results and holding the correspondence with the classical theorem at the high-temperature regime \cite{art2, art4, art5}. Due to the recent exploration of this area, the works about it are very homogeneous and based on the same conceptual and mathematical framework. 

\vspace{1.2mm}

\noindent In this article a novel approach is implemented; namely, the derivation of an analog of the classical equipartition theorem in the phase-space of quantum mechanics through the Wigner representation. First, a review and certain mathematical manipulations are made to the classical statement of the theorem so that with the aid of some results in the phase-space formulation, the version of the theorem is shown straightforward in a completely analog manner. The results derived are tested and validated, applying them to the harmonic oscillator in both high and low-temperature regimes for the particular case of  weak coupling limit.

\vspace{5mm}

\section{Classical Version of the Theorem}

Consider a system composed of $N$ particles and described by the set of generalized coordinates \{$q_i$, $p_i$\} where $i = 1,2,...,fN$ with $f$ being the number of degrees of freedom per particle.  The classical Hamiltonian of the system is $H_S(q,p)$ where $q$ and $p$ represent all the generalized coordinates, so the 3D energy equipartition theorem reads as follows \cite{Greiner1995}
\begin{equation}
\label{equipx}
    \left<q_i \frac{\partial H_S}{\partial q_i} \right> = \frac{1}{h^{3N}}\int d^{N}\mathbf{q} \hspace{0.5 mm}d^{N}\mathbf{p}\hspace{1.2 mm} \rho(\mathbf{q},\mathbf{p}) \hspace{0.5 mm}q_i \frac{\partial H_S}{\partial q_i} = \frac{1}{\beta}
\end{equation}

\begin{equation}
\label{equipp}
    \left<p_i \frac{\partial H_S}{\partial p_i} \right> = \frac{1}{h^{3N}}\int d^{N}\mathbf{q} \hspace{0.5 mm}d^{N}\mathbf{p}\hspace{1.2 mm} \rho(\mathbf{q},\mathbf{p}) \hspace{0.5 mm} p_i \frac{\partial H_S}{\partial p_i} = \frac{1}{\beta}
\end{equation}

\noindent here $h$ is the Planck constant, $\beta=(k_BT)^{-1}$ where $k_B$ is the Boltzmann constant,  $\mathbf{q}$ and $\mathbf{p}$ denote all the coordinates per particle and $\rho(\mathbf{q},\mathbf{p})$ is the corresponding phase-space density in any of the Gibbs ensembles \cite{Greiner1995}. Notice that the coordinates $q_i$ and $p_i$ have not been merged into a single coordinate $x_i$ as is customary in the demonstration of this theorem, this has a purpose as will be shortly shown. In the particular case in which the density distribution is given by the canonical ensemble, $\rho(\mathbf{q},\mathbf{p})=\exp(-\beta  H_S(q,p))/Z$ where explicitly the partition function is $Z=\frac{1}{h^{3N}} \int d^{N}\mathbf{q} \hspace{0.5 mm}d^{N}\mathbf{p}\hspace{1.2 mm}\exp-(\beta  H_S(q,p))$. It is possible, for simplicity, to think in an alternative version of this theorem for the case in which $\beta$ does not correspond with the known usual expression, but with a modified function of temperature $\beta_{mod} = \beta_{mod}(T)$, then, equations \ref{equipx} and \ref{equipp} read as follows:

\begin{equation}
\label{equipxm}
    \left<q_i \frac{\partial H_S}{\partial q_i} \right> =  \frac{1}{\beta_{mod}}
\end{equation}

\begin{equation}
\label{equippm}
    \left<p_i \frac{\partial H_S}{\partial p_i} \right> =  \frac{1}{\beta_{mod}}
\end{equation}

\noindent where now the density distribution is given by $\rho_{mod}(\mathbf{q},\mathbf{p})=
\text{exp}(-\beta_{mod}  H_S(\mathbf{q},\mathbf{p}))/Z_{mod}$.

\noindent Let now assume that the Hamiltonian of the system can be separated in two functions according to:
\begin{equation}
\label{ham}
    H_S(q,p) = F(q) + G(p),
\end{equation}
which is usually the case, so we will be able to transform equations \ref{equipx} and \ref{equipp} into more favorable forms for the connection with the quantum mechanical phase-space formulation. This is a feasible assumption since many physical systems hold it. Let's now consider the following Hamilton equations 
\begin{equation}
\label{h1}
    \dot{q}_i = \frac{\partial H_S}{\partial p_i} = \{q_i, H_S\}_{PB}
\end{equation}
\begin{equation}
\label{h2}
    \dot{p}_i = -\frac{\partial H_S}{\partial q_i} = \{p_i, H_S\}_{PB}
\end{equation}
where $\{ , \}_{PB}$ is the Poisson bracket defined for quantities $A(q,p)$ and $B(q,p)$ as 
\begin{equation}
\label{poiss}
    \{A, B\}_{PB} = \sum_i \left(\frac{\partial A}{\partial q_i}\frac{\partial B}{\partial p_i}-\frac{\partial A}{\partial p_i}\frac{\partial B}{\partial q_i} \right)
\end{equation}
Including a third function $C(q,p)$, the following identity can be demonstrated 
\begin{equation}
\label{ident}
    \{AB, C\}_{PB} = A\{B, C\}_{PB} + \{A, C\}_{PB}B
\end{equation}

\noindent Applying these results for the quantities $q_i\frac{\partial H_S}{\partial q_i}$ and $p_i\frac{\partial H_S}{\partial p_i}$, using the equations \ref{ham}, \ref{h1}, \ref{h2} and identity \ref{ident}, it can be shown that 
\begin{equation}
\label{qdH}
    q_i\frac{\partial H_S}{\partial q_i} = -q_i\{p_i, H_S\}_{PB} = -q_i\{p_i, F(q)\}_{PB} =  -\{q_i p_i, F(q)\}_{PB}
\end{equation}

\begin{equation}
\label{pdH}
    p_i\frac{\partial H_S}{\partial p_i} = p_i\{q_i, H_S\}_{PB} = p_i\{q_i, G(p)\}_{PB} =  \{p_i q_i, G(p)\}_{PB}
\end{equation}
Replacing equations \ref{qdH} and \ref{pdH} in \ref{equipxm} and \ref{equippm} respectively one obtains

\begin{equation}
\label{equipx1}
     \left<q_i \frac{\partial H_S}{\partial q_i} \right> = - \int d^{N}\mathbf{q} \hspace{0.2 mm}d^{N}\mathbf{p}\hspace{0.2mm} \frac{\exp(-\beta_{mod} H_{S}(\mathbf{q},\mathbf{p}))}{Z_{mod}}  \hspace{0.2 mm}\{q_i p_i, F(q)\}_{PB} = \frac{1}{\beta_{mod}}
\end{equation}

\begin{equation}
\label{equipp1}
      \left<p_i \frac{\partial H_S}{\partial p_i} \right> = \int d^{N}\mathbf{q} \hspace{0.2 mm}d^{N}\mathbf{p}\hspace{0.2 mm} \frac{\exp(-\beta_{mod} H_{S}(\mathbf{q},\mathbf{p}))}{Z_{mod}}  \hspace{0.2 mm} \{p_i q_i, G(p)\}_{PB} = \frac{1}{\beta_{mod}}
\end{equation}

\noindent These are the algebraic forms of the energy modified equipartition theorem that will allow us henceforth to easily connect with the quantum mechanical phase-space formulation. In the case when $\beta_{mod}=\beta$, we recover the usual equipartition theorem.

\section{Density Operator in Quantum Mechanics}

It's wellFor a non-factorizing initial state is known that the total state of a system (S) and an environment (E) in quantum mechanics can be described by the thermal equilibrium state through the density operator \cite{Gardiner2000}

\begin{equation}\label{eq:d1}
\hat{\rho}={\text{exp}(-\beta \hat{H})}/{Z}
\end{equation}
\noindent where $\hat{H}$ is the total Hamiltonian of the system plus environment acting on the total Hilbert space $\mathbb{H}=\mathbb{H_{S}}\otimes \mathbb{H_{E}}$ and $Z$ the total partition function. Explicitly, the total Hamiltonian is $\hat{H}=\hat{H}_{S}+\hat{H}_{E}+\hat{V}$, where $\hat{H}_{S}$, $\hat{H}_{E}$ and $\hat{V}$ represent the Hamiltonian of the system,  the environment and the mutual interaction respectively. The interaction $\hat{V}$ can always be written in a diagonal decomposition of a set of operators $\left \{ S\hat{S}_{\alpha} \right \}$ acting only on the system  and a set $\left \{ B\hat{B}_{\alpha} \right \}$ acting only on environment, such that $\hat{V}= \sum_{\alpha} \hat{S}_{\alpha} \otimes \hat{B}_{\alpha} $ \cite{2007a}. The description of the system of interest is given by the reduced density matrix, this object contains all the information that can be extracted by an observer of the system  \cite{2007a}. It can be shown that the reduced density matrix at second order can be written as \cite{Pachon2019} 
\begin{equation}\label{eq:d2}
\hat{\rho}_{s} \propto e^{-\beta \hat{H}_{S}} \left[1+\frac{1}{\hbar} \sum_{\alpha}  \int_{0}^{\hbar \beta} d\tau_{1}\int_{0}^{\tau_{1}} d\tau_{2} \hat{S}_{\alpha}(-i\tau_{1})\hat{S}_{\alpha}(-i\tau_{2})k_{\alpha}(\tau_1-\tau_2)\right]
\end{equation}
\noindent where $\hbar k_{\alpha}(\tau)=\left<\hat{B}_\alpha(-i\tau)\hat{B}_\alpha(0)\right>_{B}$ is the two-time correlation function of the environment operators. Equation \ref{eq:d2} shows that in general, it is not possible to describe the influence of the bath in a system by the Gibbs state as in the case of classical mechanics \cite{Pachon2019}. However, in the limit of high temperature $\hbar \beta \rightarrow 0$, independently of the interaction form of $\hat{V}$, tIn order to give a simple illustration in section 6, we consider the weak coupling limit. The reduced density matrix in this case is the Gibbs state given by

\begin{equation}\label{part}
\hat{\rho}_{S}=\text{exp}(-\beta \hat{H}_{S})/{Z_{S}}
\end{equation}

\noindent where in this caseWith $Z_{S}=\text{Tr}_{S}(\text{exp}(-\beta \hat{H}_{S}))$ being the canonical partition function.
\noindent On the other hand, it is useful to note that there is a relationship between the propagator $K(q_{f},t,q_{i},0)= \left \langle q_{f}|\text{exp}(-\frac{it}{\hbar}\hat{H}_{S}) | q_{i}  \right \rangle $ formulated in terms of path integrals \cite{2002} and the matrix elements in the position basis of equation \ref{part}. Such a  relation is evident when we replace $t\rightarrow -i\hbar \beta$ in the propagator, i.e,  $K(q_{f}, -i\hbar \beta,q_{i},0)= \left \langle q_{f}|\text{exp}(-\beta \hat{H}_{S}) | q_{i}  \right \rangle $, then one obtains

\begin{equation}\label{eq:d4}
\left \langle q_{f}| \hat{\rho}_{S}| q_{i}  \right \rangle= \frac{1}{Z_{S}} \left \langle q_{f}|\text{exp}(-\beta \hat{H}_{S}) | q_{i}  \right \rangle= \frac{1}{Z_{S}}  \int \mathfrak{D}q \hspace{1 mm} \text{exp} (-\frac{1}{\hbar}S^{E}[q])
\end{equation}

\noindent Here the integral is a functional integral running over all the functions satisfying the boundary conditions $q(0)=q_i$, $q(\hbar \beta )=q_{f}$ and  $S^{E}[q]=\int_{0}^{\hbar \beta} ds H_{S}(q(s),p(s))$ is the action of the system.
\section{Phase-Space Formulation of Quantum Mechanics}

There are different approaches to phase-space formulation of quantum mechanics that try to associate a phase-space distribution function, for example the Glauber-Sudarshan, Kirkwood, Husimi $Q-$representations \cite{phasespace, phasespace2}, etc. There is a particular approach that consists in associating every quantum observable $\hat{O}(\hat{q},\hat{p})$ to a phase-space function $O_W(\mathfrak{q},\mathfrak{p})$ (also called symbol) by means of a bijection $\Phi$, called the Weyl transformation and defined as \cite{Wezeman2014WeylQA}
\begin{equation}
\label{weyltr}
\Phi(\hat{O}(\hat{q},\hat{p})) = O_{W}(\mathbf{\mathfrak{q}},\mathbf{\mathfrak{p}})= \int  d^{N}\mathfrak{u} \hspace{1.2 mm}  \exp(-\frac{i}{\hbar} \mathbf{\mathfrak{p}} \cdot \mathbf{\mathfrak{u}}) \left \langle \mathbf{\mathfrak{q}}+\frac{\mathbf{\mathfrak{u}}}{2}\Big|\hat{O}(\hat{q},\hat{p}) \Big|\mathbf{\mathfrak{q}}-\frac{\mathbf{\mathfrak{u}}}{2} \right \rangle
\end{equation}
where 
\begin{equation}
\label{weyltrq}
\Phi(\hat{q}) = \mathfrak{q}
\end{equation}
\begin{equation}
\label{weyltrp}
\Phi(\hat{p}) =\mathbf{\mathfrak{p}}
\end{equation}
are the corresponding symbols of position and momentum operators. The Wigner distribution $W(\mathbf{\mathfrak{q}},\mathbf{\mathfrak{p}})$ defined as the Weyl symbol of the density operator $\hat{\rho}$ by 
\begin{equation}
\label{wigner}
W(\mathbf{\mathfrak{q}},\mathbf{\mathfrak{p}})= \rho_{W}(\mathfrak{q},\mathfrak{p})= \frac{1}{(2 \pi \hbar )^{f}}\int  d^{N}\mathfrak{u} \hspace{1.2 mm}  \exp(-\frac{i}{\hbar} \mathbf{\mathfrak{p}} \cdot \mathbf{\mathfrak{u}}) \left \langle \mathbf{\mathfrak{q}}+\frac{\mathbf{\mathfrak{u}}}{2}\Big|\hat{\rho}(\hat{q},\hat{p}) \Big|\mathbf{\mathfrak{q}}-\frac{\mathbf{\mathfrak{u}}}{2} \right \rangle
\end{equation}
is the corresponding phase-space distribution. This allows to compute averages in a similar way as in classical mechanics:
\begin{equation}
\label{P2}
\left \langle \hat{O} \right \rangle=\int d^{N}\mathfrak{q} \hspace{0.5 mm}d^{N}\mathfrak{p}\hspace{1.2 mm} W(\mathfrak{q},\mathfrak{p})  O_{W}. 
\end{equation}

\noindent Since oOur purpose is to create an analogy between the classical theorem and its counterpart in phase-space quantum mechanics, we notice that in the particular case where we analyze the energetic distribution of the system. Thus the Hamiltonian that appears in the classical version of the energy equipartition theorem is the Hamiltonian  of the system $\hat{H}_S$. Therefore, it is of interest to calculate the same average in quantum mechanics formulated in phase-space in order to create an analogy. To do that, we consider $\hat{q_{i}}=\hat{q_{i}}\otimes \mathbb{I}$, $\hat{p_{i}}=\hat{p_{i}}\otimes \mathbb{I}$  and $\frac{\partial \hat{H}_{s}}{\partial \hat{q}_{i}}= \frac{\partial \hat{H}_{s}}{\partial \hat{q}_{i}} \otimes \mathbb{I}$, $\frac{\partial \hat{H}_{s}}{\partial \hat{p}_{i}}= \frac{\partial \hat{H}_{s}}{\partial \hat{p}_{i}} \otimes \mathbb{I}$, so we have
\begin{equation}
\label{qequipq}
    \left<\hat{q_i} \frac{\partial \hat{H}_S}{\partial \hat{q}_i}\otimes \mathbb{I}  \right> = \text{Tr}_{S}\left(\hat{\rho}_{S} \hat{q_i} \frac{\partial \hat{H}_S}{\partial \hat{q}_i}\right)=\int d^{N}\mathbf{\mathfrak{q}} \hspace{0.5 mm}d^{N}\mathbf{\mathfrak{p}}\hspace{1.2 mm} W_{s}(\mathbf{\mathfrak{q}},\mathbf{\mathfrak{p}}) \left( \hat{q_i} \frac{\partial \hat{H}_S}{\partial \hat{q}_i}\right)_{W}\hspace{0.5 mm} 
\end{equation}
\begin{equation}
\label{qequipp}
    \left<\hat{p_i} \frac{\partial \hat{H}_S}{\partial \hat{p}_i}\otimes \mathbb{I}  \right> = \text{Tr}_{S}\left(\hat{\rho}_{S}  \hat{p_i} \frac{\partial \hat{H}_S}{\partial \hat{p}_i}\right)=\int d^{N}\mathbf{\mathfrak{q}} \hspace{0.5 mm}d^{N}\mathbf{\mathfrak{p}}\hspace{1.2 mm} W_{s}(\mathbf{\mathfrak{q}},\mathbf{\mathfrak{p}}) \left( \hat{p_i} \frac{\partial \hat{H}_S}{\partial \hat{p}_i}\right)_{W}\hspace{0.5 mm} 
\end{equation}

\noindent Where $W_{s}(\mathbf{\mathfrak{q}},\mathbf{\mathfrak{p}})$ is the Wigner distribution associated with the reduced density matrix of the system. 

\noindent A Weyl symbol of particular interest is the one associated to the product of two operators given by \cite{Wezeman2014WeylQA}
\begin{equation}
\label{moyalprod}
\begin{split}
    \Phi(\hat{A}(\hat{q},\hat{p})\hat{B}(\hat{q},\hat{p})) = A_{W}(\mathbf{\mathfrak{q}},\mathbf{\mathfrak{p}})\star_M B_{W}(\mathbf{\mathfrak{q}},\mathbf{\mathfrak{p}}) \\= A_{W}(\mathbf{\mathfrak{q}},\mathbf{\mathfrak{p}})\text{exp}\left(\frac{i\hbar}{2}\left(\overleftarrow{\partial_\mathfrak{q}}\overrightarrow{\partial_\mathfrak{p}}-\overleftarrow{\partial_\mathfrak{p}}\overrightarrow{\partial_\mathfrak{q}}\right)\right)B_{W}(\mathbf{\mathfrak{q}},\mathbf{\mathfrak{p}})
\end{split}
\end{equation}
where $\overleftarrow{\partial_\mathfrak{q}}$ denotes the derivative with respect to $\mathfrak{q}$ acting to the left and similarly for the other derivatives. The $\star_M$ is called the Moyal product and allows one to define the associated symbol for the commutator $\{,\}_M$ known as Moyal bracket  
\begin{equation}
\label{moyalbracket}
\begin{split}
    \Phi([\hat{A}(\hat{q},\hat{p}),\hat{B}(\hat{q},\hat{p})]) = \{A_{W}(\mathbf{\mathfrak{q}},\mathbf{\mathfrak{p}}), B_{W}(\mathbf{\mathfrak{q}},\mathbf{\mathfrak{p}})\}_M \\= \frac{2}{\hbar} A_{W}(\mathbf{\mathfrak{q}},\mathbf{\mathfrak{p}})\text{sin}\left(\frac{\hbar}{2}\left(\overleftarrow{\partial_\mathfrak{q}}\overrightarrow{\partial_\mathfrak{p}}-\overleftarrow{\partial_\mathfrak{p}}\overrightarrow{\partial_\mathfrak{q}}\right)\right)B_{W}(\mathbf{\mathfrak{q}},\mathbf{\mathfrak{p}})
\end{split}
\end{equation}
which expanded is expressed as 
\begin{equation}
\label{moyalbracketexpanded}
\begin{split}
    \{A_{W}(\mathbf{\mathfrak{q}},\mathbf{\mathfrak{p}}), B_{W}(\mathbf{\mathfrak{q}},\mathbf{\mathfrak{p}})\}_M = 2 \sum_{s=0}^\infty\frac{(-1)^s}{(2s+1)!}\left( \frac{-\hbar}{2}\right)^{2s}\sum_{t=0}^{2s+1}\frac{(-1)^t(2s+1)!}{(2s+1-t)!t!}\\
    \times\left[ \frac{\partial^t}{\partial \mathfrak{q}^t}  \frac{\partial^{2s+1-t}A_{W}}{\partial \mathfrak{p}^{2s+1-t}} \right]\left[  \frac{\partial^{2s+1-t}}{\partial \mathfrak{q}^{2s+1-t}} \frac{\partial^t B_{W}}{\partial \mathfrak{p}^t}\right]\\ \\
    = \{A_{W}(\mathbf{\mathfrak{q}},\mathbf{\mathfrak{p}}), B_{W}(\mathbf{\mathfrak{q}},\mathbf{\mathfrak{p}})\}_{PB} + \mathcal{O}(\hbar^2)
\end{split}
\end{equation}

\section{Version of the Theorem in Phase-Space
Formulation of Quantum Mechanics}

Proceeding in a similar way as in section 1, by assuming 
\begin{equation}
\label{qham}
    \hat{H}_S(\hat{q},\hat{p}) = \hat{F}(\hat{q}) + \hat{G}(\hat{p}),
\end{equation}
and considering the commutator for operators $\hat{A}(\hat{q},\hat{p})$ and $\hat{B}(\hat{q},\hat{p})$ defined as
\begin{equation}
\label{commut}
    [\hat{A}, \hat{B}] = \hat{A}\hat{B} - \hat{B}\hat{A}
\end{equation}
the following identities can be demonstrated \cite{Sakurai2020}
 \begin{equation}
\label{qid1}
[\hat{A}\hat{B}, \hat{C}] = \hat{A}[\hat{B}, \hat{C}] + [\hat{A}, \hat{C}]\hat{B}
\end{equation}
\begin{equation}
\label{qid2}
[\hat{q}_i,\mathfrak{G}(\hat{p})] = i\hbar \frac{d \mathfrak{G}(\hat{p})}{d \hat{p}_i}
\end{equation}
\begin{equation}
\label{qid3}
[\hat{p}_i,\mathfrak{F}(\hat{q})] = - i\hbar \frac{d \mathfrak{F}(\hat{q})}{d \hat{q}_i}
\end{equation}

Where $\mathfrak{F}$ and $\mathfrak{G}$ are arbitrary functions of operators $\hat{q}$ and $\hat{p}$. Thus, using equations \ref{qid2} and \ref{qid3}, the quantum mechanical analog of Hamilton equations can be stated as 
\begin{equation}
    \label{hamiltq}
    \frac{d\hat{q}_i}{dt} = \frac{\partial \hat{H}_S}{d\hat{p}_i}  =  \frac{1}{i\hbar}[\hat{q}_i, \hat{H}_S] 
\end{equation}
\begin{equation}
    \label{hamiltp}
    \frac{d\hat{p}_i}{dt} = -\frac{\partial \hat{H}_S}{d\hat{q}_i}  =  \frac{1}{i\hbar}[\hat{p}_i, \hat{H}_S] 
\end{equation}
Similarly one arrives at
\begin{equation}
\label{qqdH}
    \hat{q}_i\frac{\partial \hat{H}_S}{\partial \hat{q}_i} = -\frac{\hat{q}_i}{i\hbar}[\hat{p}_i, \hat{H}_S] = -\frac{\hat{q}_i}{i\hbar}[\hat{p}_i, \hat{F}(\hat{q})] =  -\frac{1}{i\hbar}[\hat{q}_i\hat{p}_i, \hat{F}(\hat{q})]
\end{equation}

\begin{equation}
\label{qpdH}
    \hat{p}_i\frac{\partial \hat{H}_S}{\partial \hat{p}_i} = \frac{\hat{p}_i}{i\hbar}[\hat{q}_i, \hat{H}_S] = \frac{\hat{p}_i}{i\hbar}[\hat{q}_i, \hat{G}(\hat{p})] =  \frac{1}{i\hbar}[\hat{p}_i\hat{q}_i, \hat{G}(\hat{p})]
\end{equation}
By replacing in equations \ref{qequipq} and \ref{qequipp} the result is 
\begin{equation}
\label{qequipqmoyal}
\begin{split}
\left<\hat{q_i} \frac{\partial \hat{H}_S}{\partial \hat{q}_i}\otimes \mathbb{I}  \right> =  - \int d^{N}\mathbf{\mathfrak{q}} \hspace{0.5 mm}d^{N}\mathbf{\mathfrak{p}}\hspace{1.2 mm} W_{s}(\mathbf{\mathfrak{q}},\mathbf{\mathfrak{p}}) \left( \frac{1}{i\hbar}[\hat{q}_i\hat{p}_i, \hat{F}(\hat{q})]\right)_{W} \\
= - \int d^{N}\mathbf{\mathfrak{q}} \hspace{0.5 mm}d^{N}\mathbf{\mathfrak{p}}\hspace{1.2 mm} W_{s}(\mathbf{\mathfrak{q}},\mathbf{\mathfrak{p}}) \{ \mathfrak{q}_i \mathfrak{p}_i, F(\mathfrak{q})\}_{M}
\end{split}
\end{equation}

\begin{equation}
\label{qequippmoyal}
\begin{split}
\left<\hat{p_i} \frac{\partial \hat{H}_S}{\partial \hat{p}_i}\otimes \mathbb{I}  \right> =  \int d^{N}\mathbf{\mathfrak{q}} \hspace{0.5 mm}d^{N}\mathbf{\mathfrak{p}}\hspace{1.2 mm} W_{s}(\mathbf{\mathfrak{q}},\mathbf{\mathfrak{p}}) \left( \frac{1}{i\hbar}[\hat{p}_i\hat{q}_i, \hat{G}(\hat{p})]\right)_{W} \\
= \int d^{N}\mathbf{\mathfrak{q}} \hspace{0.5 mm}d^{N}\mathbf{\mathfrak{p}}\hspace{1.2 mm} W_{s}(\mathbf{\mathfrak{q}},\mathbf{\mathfrak{p}}) \{ \mathfrak{p}_i \mathfrak{q}_i, G(\mathfrak{p})\}_{M}
\end{split}
\end{equation}
Expanding the Moyal bracket as in \ref{moyalbracketexpanded} and noticing that the $O(\hbar^2)$ vanish since derivatives of third-order or higher cancel out one arrives to
\begin{equation}
\label{equipqexpansion}
\begin{split}
\left<\hat{q_i} \frac{\partial \hat{H}_S}{\partial \hat{q}_i}\otimes \mathbb{I}  \right> = - \int d^{N}\mathbf{\mathfrak{q}} \hspace{0.5 mm}d^{N}\mathbf{\mathfrak{p}}\hspace{1.2 mm} W_{s}(\mathbf{\mathfrak{q}},\mathbf{\mathfrak{p}}) \{ \mathfrak{q}_i \mathfrak{p}_i, F(\mathfrak{q})\}_{M}\\
= - \int d^{N}\mathbf{\mathfrak{q}} \hspace{0.5 mm}d^{N}\mathbf{\mathfrak{p}}\hspace{1.2 mm} W_{s}(\mathbf{\mathfrak{q}},\mathbf{\mathfrak{p}}) \{ \mathfrak{q}_i \mathfrak{p}_i, F(\mathfrak{q})\}_{PB} \\ 
\end{split}
\end{equation}

\begin{equation}
\label{equippexpansion}
\begin{split}
\left<\hat{p_i} \frac{\partial \hat{H}_S}{\partial \hat{p}_i}\otimes \mathbb{I}  \right> = \int d^{N}\mathbf{\mathfrak{q}} \hspace{0.5 mm}d^{N}\mathbf{\mathfrak{p}}\hspace{1.2 mm} W_{s}(\mathbf{\mathfrak{q}},\mathbf{\mathfrak{p}}) \{ \mathfrak{p}_i \mathfrak{q}_i, G(\mathfrak{p})\}_{M}\\
= \int d^{N}\mathbf{\mathfrak{q}} \hspace{0.5 mm}d^{N}\mathbf{\mathfrak{p}}\hspace{1.2 mm} W_{s}(\mathbf{\mathfrak{q}},\mathbf{\mathfrak{p}}) \{ \mathfrak{p}_i \mathfrak{q}_i, G(\mathfrak{p})\}_{PB}
\end{split}
\end{equation}

\noindent TAs we will show in section 7, these equations \ref{equipqexpansion} and \ref{equippexpansion} constitute the energy equipartition theorem for the system in the phase-space formulation of quantum mechanics as an analogy to the formulation in classical mechanics in the weak coupling limit as a consequence of the high-temperature regime. Notice the explicitly the resemblance with the classical version as shown in equations \ref{equipx} and \ref{equipp}. The second terms of equations \ref{qequipqmoyal} and \ref{qequippmoyal} vanish since derivatives of third-order or higher cancel out.

\section{Application to the Quantum Harmonic Oscillator}

As an illustration, let us consider the system to be a harmonic oscillator in the high and low-temperature regime. In the first case, according to equation \ref{eq:d2}, we do not need to specify the interaction of the environment-system.Exactly the same result can be obtained for the environment, but with the Wigner distribution function associated to the reduced density matrix of the bath.

\section{Application to the Quantum Harmonic Oscillator}


As an illustrative example of the proposed method in the article for calculating the energy, let us consider the system to be a harmonic oscillator in the high and low-temperature regimes.  First, we calculate the Wigner distribution associated with the reduced density matrix: in the weak coupling limit
\begin{equation}
\label{E1}
W_{s}(\mathbf{\mathfrak{q}},\mathbf{\mathfrak{p}})=\frac{1}{Z_{S}} \frac{1}{2 \pi \hbar} \int  du \hspace{1.2 mm}  \exp(-\frac{i}{\hbar} \mathbf{\mathfrak{p}} \cdot \mathbf{\mathfrak{u}}) \left \langle \mathbf{\mathfrak{q}}+\frac{\mathbf{\mathfrak{u}}}{2}|\exp(-\beta \hat{H}_S) |\mathbf{\mathfrak{q}}-\frac{\mathbf{\mathfrak{u}}}{2} \right \rangle
\end{equation}

\noindent with $\hat{H}_{S}= \frac{1}{2m} \hat{p}^2+ \frac{1}{2}m \omega^2\hat{q}^2$. As we said in section 2, the matrix elements of the density operator involved in the integral can be calculated in the imaginary time path integral formalism, i.e, we use the expression of the propagator of a harmonic oscillator and we replace $t=-i\hbar \beta$. We do this in the expression provided by \cite{Sakurai2020} to get $\left \langle \mathbf{\mathfrak{q}}+\frac{\mathbf{{\mathfrak{u}}}}{2}\Big|\text{exp}(-\beta \hat{H}_{S})\Big|\mathbf{{\mathfrak{q}}}-\frac{\mathbf{{\mathfrak{u}}}}{2} \right \rangle=\sqrt{\frac{m \omega}{2 \pi \hbar \text{sinh}(\hbar \beta \omega)}}\text{exp}\big(-\frac{m \omega}{\hbar}q^{2} \text{tanh}(\frac{\hbar \omega \beta}{2})-\frac{m \omega}{4\hbar}\mathbf{\mathfrak{u}}^{2} \text{coth}(\frac{\hbar \omega \beta}{2}))$ and $Z_{S}= (2 \text{sinh}(\frac{\hbar \omega \beta}{2}))^{-1}$. Finally, we replace this in equation \ref{E1} and get
\begin{equation}
\begin{split}
\label{E2}
W_{s}(\mathbf{\mathfrak{q}},\mathbf{\mathfrak{p}})=\frac{1}{\pi \hbar} \tanh\left(\frac{\hbar \omega \beta}{2}\right) \text{exp} \left(-\frac{\tanh(\frac{\hbar \omega \beta}{2})}{\omega \hbar}\left(\frac{\mathbf{\mathfrak{p}}^2}{m}+m \omega^2 \mathbf{\mathfrak{q}}^2\right)\right)
\end{split}
\end{equation}

\noindent because we have used the result of the reduced density matrix in the high-temperature limit, we need to impose the same limit in equation \ref{E2}. In that case, we find that the Wigner distribution function is just the reduced density matrix of the system when we replace the operators $\hat{p}$ and $\hat{q}$ for their corresponding symbols $p$ and $q$In the high-temperature regime, we find that the Wigner distribution function is just the reduced density matrix of the system, i.e the classical phase-space density in the canonical ensemble of a harmonic oscillator,

\begin{equation}
\begin{split}
\label{E3}
W_{s}(\mathbf{\mathfrak{q}},\mathbf{\mathfrak{p}})=\frac{1}{Z} \text{exp} \left(-\beta \left(\frac{\mathbf{\mathfrak{p}}^2}{2m}+\frac{m \omega^2}{2} \mathbf{\mathfrak{q}}^2\right)\right)=\frac{\text{exp}\left(-\beta H_{S}(\mathbf{\mathfrak{q}},\mathbf{\mathfrak{p}})\right)}{Z}
\end{split}
\end{equation}

Where $Z = \frac{2\pi}{\beta \omega}$. \noindent Finally, because  equation \ref{E3} corresponds exactly with the probability density function for the canonical ensemble as in the case of classical mechanics, equations \ref{equipqexpansion} and \ref{equippexpansion} are equal to $k_{B} T$ using equations \ref{equipx1} and \ref{equipp1}.

\noindent In the low -temperature regime, formally the description of the system is given by equation \ref{eq:d2} and strictly speaking the equation \ref{E2} is no longer valid. However, if we suppose an interaction between the environment and the system such that $\left [ \hat{H}_{S},\hat{V} \right ]=0$, the system is even well described by the reduced density matrix as in the high temperature regime \cite{Pachon2019} and therefore equation \ref{E2} is valid. Then,

\begin{equation}
\label{osl}
\begin{split}
\left<\hat{p_i} \frac{\partial \hat{H}_S}{\partial \hat{p}_i}\otimes \mathbb{I}  \right> = \lim_{\beta \rightarrow \infty} \int d\mathbf{\mathfrak{q}} \hspace{0.5 mm}d\mathbf{\mathfrak{p}}\hspace{1.2 mm} W_{s}(\mathbf{\mathfrak{q}},\mathbf{\mathfrak{p}}) \{ \mathfrak{p}_i \mathfrak{q}_i, G(\mathfrak{p})\}_{M}\\
= \int d\mathbf{\mathfrak{q}} \hspace{0.5 mm}d\mathbf{\mathfrak{p}}\hspace{1.2 mm} 
\frac{1}{\pi \hbar} \exp \left(-\frac{2 H_{S}(\mathbf{\mathfrak{q}},\mathbf{\mathfrak{p}}) }{\omega \hbar} \right)\{ \mathfrak{p}_i \mathfrak{q}_i, G(\mathfrak{p})\}_{PB}=\frac{\hbar \omega}{2}
\end{split}
\end{equation}

\noindent where in the last equality we have used equation \ref{equipp1} and where
$Z_{mod}= \int d \mathbf{q} \hspace{0.5 mm}d \mathbf{p}\hspace{1.2 mm} \exp(-\beta_{mod}  H_{S}(\mathbf{\mathfrak{q}},\mathbf{\mathfrak{p}q},\mathbf{p}))=\pi \hbar$, and $\beta_{mod}=\frac{2}{\omega \hbar}$.

This simple example shows for this particular case, that if one expects to create a true correspondence between the classical equipartition theorem discussed in section 1, and the theorem in quantum mechanics phase-space formulation, it must be done in the high-temperature regime because in the low-temperature regime one gets the modified density distribution. Further, for a model whose coordinates are bilinearly coupled as in \cite{Bialas2018}, there are found explicit expressions for equations \ref{qequipq} and \ref{qequipp} in both regimes for the weak coupling limit, there it is shown that the energy is not equally distributed for the damped harmonic oscillator. This can also be seen from \cite{Pachon2010} where the Wigner distribution function for equations \ref{equippexpansion} and \ref{equipqexpansion} does not correspond with the Gibbs distribution.

\section{Theorem in the Classical Limit}

Motivated by this example, we expect that in the classical limit the Wigner function described by equation \ref{E1} behaves as the density probability function in the canonical ensemble for any system. Then,In the general case of non-factorizing initial conditions in the high-temperature regime equation, \ref{eq:d1} takes the form of Gibbs state and it can be shown \cite{Yan2018} that in this limit the Wigner distribution function coincides with the same distribution as in the classical case,

\begin{equation}\label{th0}
W_{s}(\mathbf{\mathfrak{q}},\mathbf{\mathfrak{p}})=\frac{\text{exp}(-\beta H_{S}(\mathbf{\mathfrak{q}},\mathbf{\mathfrak{p}}))}{Z}
\end{equation}

\noindent Where in this case we have used the fact that the Wigner distribution function must be normalized.
\noindent Using this in  equations  \ref{equipqexpansion} and \ref{equippexpansion}, we have

\begin{equation}
\label{th1}
\begin{split}
\left<\hat{q_i} \frac{\partial \hat{H}_S}{\partial \hat{
q}_i}\otimes \mathbb{I}  \right> \approx -\int d^{N}\mathbf{\mathfrak{q}} \hspace{0.5 mm}d^{N}\mathbf{\mathfrak{p}}\hspace{1.2 mm} \frac{\text{exp}(-\beta H_{S}(\mathbf{\mathfrak{q}},\mathbf{\mathfrak{p}}))}{Z} \{ \mathfrak{p}_i \mathfrak{q}_i, F(\mathfrak{q})\}_{PB}
\end{split}
\end{equation}

\begin{equation}
\label{th2}
\begin{split}
\left<\hat{p_i} \frac{\partial \hat{H}_S}{\partial \hat{p}_i}\otimes \mathbb{I}  \right> \approx \int d^{N}\mathbf{\mathfrak{q}} \hspace{0.5 mm}d^{N}\mathbf{\mathfrak{p}}\hspace{1.2 mm} \frac{\text{exp}(-\beta H_{S}(\mathbf{\mathfrak{q}},\mathbf{\mathfrak{p}}))}{Z} \{ \mathfrak{p}_i \mathfrak{q}_i, G(\mathfrak{p})\}_{PB}
\end{split}
\end{equation}

\noindent This is according to the classical theorem equal to $k_{B} T$ (see equation \ref{equipx1} and \ref{equipp1}). Then
\begin{equation}\label{th3}
\left<\hat{q_i} \frac{\partial \hat{H}_S}{\partial \hat{q}_i}\otimes \mathbb{I}  \right>\approx k_B T 
\end{equation}
\begin{equation}\label{th4}
\left<\hat{p_i} \frac{\partial \hat{H}_S}{\partial \hat{p}_i}\otimes \mathbb{I}  \right>\approx k_B T 
\end{equation}

\noindent These equations represent what we expect to get in the classical limit, i.e, the equipartition theorem.  It must be stressed that this result has some approximations, first
restrictions, and limitations. First, in general, the total state given by equation 
\ref{eq:d1} must contain some additional conditions resulting from experimentally achievable preparations \cite{Grabert1988}, second the result is derived based on an approximation method derived in \cite{Yan2018}, and second we approximate the product $\hbar \beta$ tends to zero to guarantee that the state is described by the Gibbs state. The latter is guaranteed formally when $T \rightarrow \infty$ and as long as the second-order perturbation theory is valid \cite{Pachon2019}. In this case, no matter what kind of interaction is described by the system and the environment, we get  \ref{th3} and \ref{th4}. On the other hand, the result is only valid when one works with separable Hamiltonians which is a particular case of more general Hamiltonians presented in the literature \cite{Nicacio_2021}.


\section*{Acknowledgments}
One of the authors (J. R.) acknowledges University of Antioquia for the exclusive dedication program. Financial support was provided by the CODI-UdeA 2020-34211 Simulmag2 and 2017-16253 projects. (E. M. and A. R.) thank the professors Leonardo Pachon and Juan David Jaramillo for enjoyable and stimulating discussions.

\bibliographystyle{unsrt}
\bibliography{refs}

\end{document}